\documentstyle[prc,aps,eqsecnum,epsfig]{revtex}            
\def\be{\begin{equation}}
\def\ee{\end{equation}}
\def\bea{\begin{eqnarray}}
\def\eea{\end{eqnarray}}

\def\g5{\gamma_5}

\def\vepjnm{\varepsilon^{njm}}

\newcommand{\noi}{\noindent}

\def\Journal#1#2#3#4{{#1}{\bf #2}, #3 (#4)}

\def\APNY{ Ann. Phys. (N.Y.)\,\,} 
\def\NPA{{ Nucl. Phys.} {\bf A}}
\def\NPB{{ Nucl. Phys.} {\bf B}}
\def\PRL{ Phys. Rev. Lett.\,\,}
\def\PRC{{ Phys. Rev.} C\,\,}
\def\PRD{{ Phys. Rev.} D\,\,}
\def\FBS{Few--Body Systems\,\,}
\def\PLB{{ Phys. Lett.} B\,\,}

\def\PR{ Phys. Rep.\,\,}
\def\CJP{ Czech. J. Phys.\,\,}
\def\FECAY{ Fiz. Elem. Chastits At. Yadra\,\,}
\def\SJPN{ Sov. J. Part. Nucl.\,\,}
\def\JPG{ J. Phys. G: Nucl. Part. Phys.\,\,}

\begin{document}

\draft 

\title{
{\large{\bf On the pion electroproduction amplitude}}}
\author{E.~Truhl\'{\i}k\footnote{Email address: truhlik@ujf.cas.cz}}
\address{Institute of Nuclear Physics, Academy of Sciences of the Czech Republic,
CZ--250 68 \v{R}e\v{z} n. Prague, Czechia
}
\maketitle

\begin{abstract}
\noi
We analyze amplitudes for the pion electroproduction on proton
derived from Lagrangians based on the local chiral $SU(2)\times SU(2)$ symmetries. 
We show that such amplitudes do contain information on the nucleon weak axial form factor
$F_A$ in both soft and hard pion regimes. This result invalidates recent Haberzettl's claim
that the pion electroproduction at threshold cannot be used to extract
any information regarding $F_A$. 
\end{abstract}


\noi


\section{Introduction}

The low energy theorems for the pion production by an external electroweak interaction were
formulated after the development of the current algebra and PCAC \cite{ADa},\cite{AFFR}.
According to them, the amplitude $M^{nj}_\lambda(q,k)$ for the production of a soft pion $\pi^n(q)$
off the nucleon by a vector--isovector current $J^j_\lambda(k)$
\be
\hat{J}^j_\lambda(k)\,+\,N(p)\,\longrightarrow\,\pi^n(q)\,+\,N(p\,')\,,  \label{reac}
\ee
given by the matrix element
\be
M^{nj}_\lambda(q,k)\,=\,\left<N(p\,')\,\pi^n(q)\,|\hat{J}^j_\lambda(0)|\,N(p)\right>\,,  \label{Mnjl}
\ee
is written as
\be
f_\pi\,M^{nj}_\lambda(q,k)\,\stackrel{q\,\rightarrow\,0}{\longrightarrow}\,i q_\mu\,
\left<p\,'|\int d^4 y e^{-iqy}\,T\left(\hat{J}^n_{5\mu}(y)\,\hat{J}^j_\lambda(0)\right)|p\right>
+\vepjnm\left<p\,'|\hat{J}^m_{5\lambda}(0)|p\right>\,.   \label{spMnjl}
\ee
Here the matrix element of the nucleon weak axial current is \cite{STK}
\be 
\left<p\,'|\hat{J}^m_{5\lambda}(0)|p\right>\,=\,i\bar{u}(p\,')\left[\,g_A F_A(q^2_1)\gamma_\lambda\g5\,-\,2igf_\pi
\Delta^\pi_F(q^2_1)q_{1\lambda}\g5\,\right]\frac{\tau^m}{2}u(p)\,,  \label{JAs}
\ee
$q_1=p\,'-p=k-q$ and $g_A=1.267$.

\begin{figure}[h!]
\centerline{
\epsfig{file=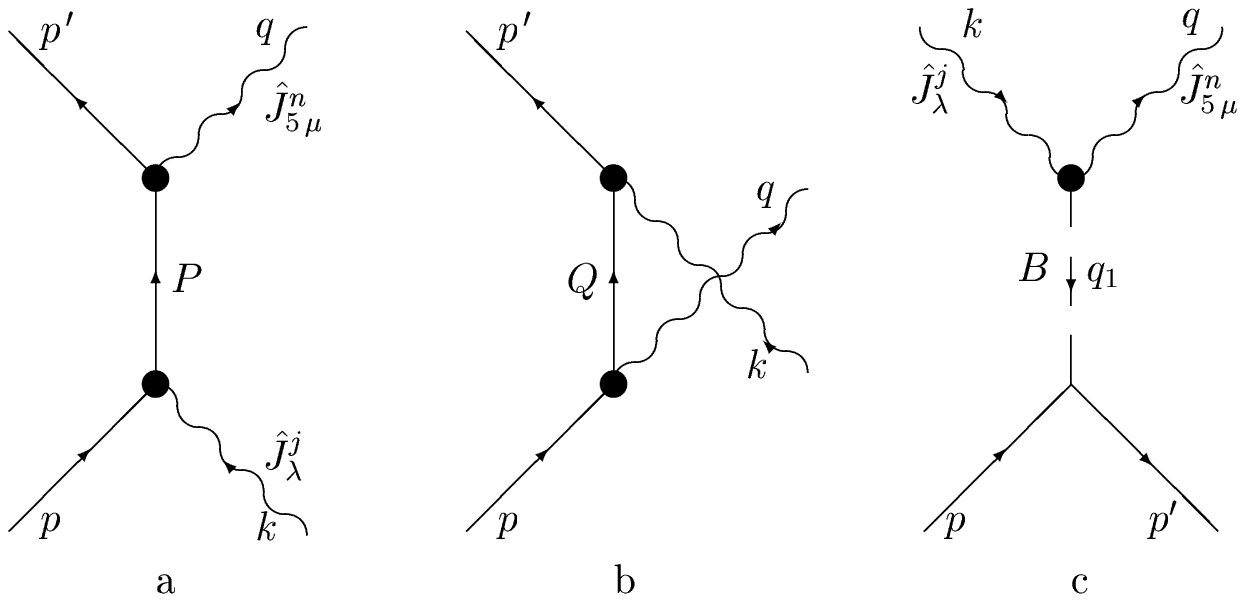}
}
\vskip 0.4cm
\caption{
The current--current amplitudes contributing to the first term on the right hand side of Eq.(\ref{spMnjl}). 
Graphs a,b -- the nucleon terms; graph c -- the contact terms of the $B\,=\,\pi$ or $a_1$ range.
}
\label{figCCA}
\end{figure}

Starting from Eq.(\ref{spMnjl}), a "master formula" for the amplitude $M^{nj}_\lambda(q,k)$ can be derived
\cite{Ad}. For this purpose, the contribution to the divergence due to the coupling of the axial current to the
external nucleon lines (Fig.\,\ref{figCCA}a and \ref{figCCA}b) can be extracted from the current--current
amplitudes. The resulting equation is
\bea 
f_\pi\,M^{nj}_\lambda(q,k)\,&=&\,\frac{gf_\pi}{2M}\,\bar{u}(p\,')\left[\not q \tau^n S_F(P) \hat{J}^j_\lambda(k)
\,+\,\hat{J}^j_\lambda(k) S_F(Q) \not q \tau^n\right]u(p) \nonumber \\
&& +\vepjnm i \bar{u}(p\,')\left[g_AF_A(q^2_1)\gamma_\lambda\g5\,-\,2igf_\pi
\Delta^\pi_F(q^2_1)q_{1\lambda}\g5\,\right]\frac{\tau^m}{2}u(p)\,.   \label{AME}
\eea
The vector--isovector current $\hat{J}^j_\lambda(k)$ is defined as
\be
\hat{J}^j_\lambda(k)\,=\,i\left[F^V_1(k)\gamma_\lambda\,-\,\frac{\kappa^V}{2M}F^V_2(k)\sigma_{\lambda \eta}
k_\eta\right]\frac{\tau^j}{2}\,.  \label{JV}
\ee

\begin{figure}[h!]
\centerline{
\epsfig{file=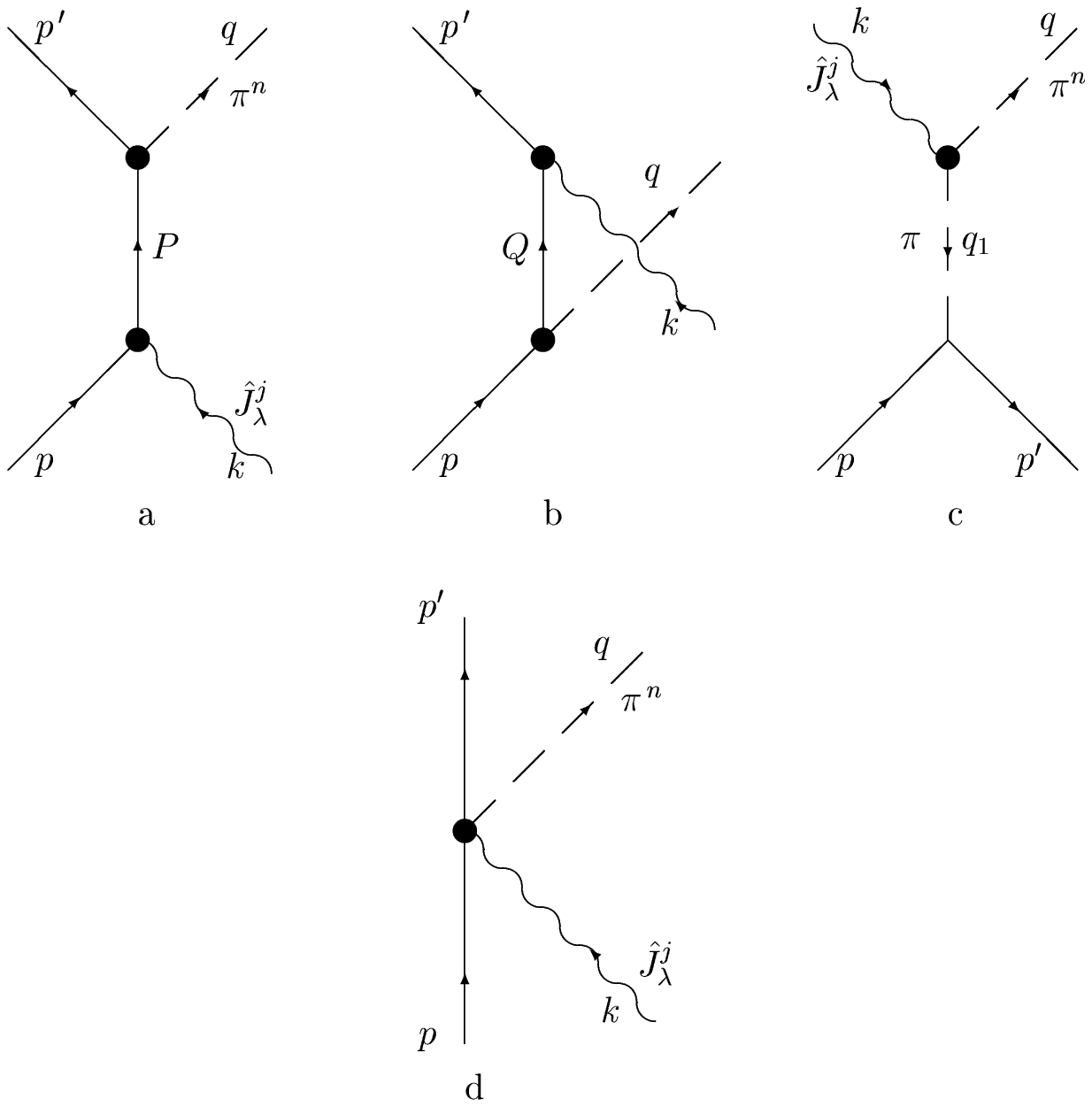}
}
\vskip 0.4cm
\caption{
The soft pion production amplitude according to Eq.(\ref{sppa}). 
Graphs a,b -- the nucleon Born terms; graph c -- the pion pole term; graph d -- the Kroll--Ruderman term.
}
\label{figSPPA}
\end{figure}

It is clear that besides the nucleon Born terms, the soft pion amplitude Eq.\,(\ref{AME}) contains the Kroll--Ruderman 
(contact) term including the nucleon weak axial form factor, and almost the pion pole production term, which is expected 
to contribute for both the pion photo-- and electroproduction. To have this term present, Adler \cite{Ad}
changed by hand $q_1\,\rightarrow\,q_1-q$ in the induced pseudoscalar term [the last term on the right hand side of
Eq.\,(\ref{AME})], thus adding an ${\cal O}(q)$ piece. It was shown \cite{ITr} that 
the evaluation of the contribution to the divergence of the current--current amplitudes using the current algebra
and PCAC due to the process of Fig.\,\ref{figCCA}c with $B=\pi$ yields two pieces \cite{BI}: one is exactly 
the pion pole production term and the other one cancels the induced pseudoscalar term. Then the soft pion production amplitude
corresponds to Fig.\,\ref{figSPPA} and it is
\bea
M^{nj}_\lambda(q,k)\,&=&\,\frac{g}{2M}\,\bar{u}(p\,')\left[\,\not q \tau^n S_F(P) \hat{J}^j_\lambda(k)
\,+\,\hat{J}^j_\lambda(k) S_F(Q) \not q \tau^n\,\right]u(p) \nonumber \\
&& +\, \frac{(k-2q)_\lambda}{(k-q)^2+m^2_\pi}\,F_\pi(k^2)\,\vepjnm\, g\,\bar{u}(p\,')\, \g5 \tau^m\, u(p)   \nonumber \\
&& +\,i\,\frac{g_A}{2f_\pi}\,F_A[(k-q)^2]\,\vepjnm\, \bar{u}(p\,')\,\gamma_\lambda\g5 \tau^m\,u(p)\,.
 \label{sppa}
\eea
The $q$ dependence in the form factor $F_A[(k-q)^2]$ can be neglected.

Elimination of the induced pseudoscalar term from the pion electroproduction amplitude and the restoration
of the pion pole production term was discussed also in Ref.\,\cite{DR} and Ref.\,\cite{DT}.
Nevertheless, the amplitude with the induced pseudoscalar term  was applied to extract
the axial and pseudoscalar form factors from the pion electroproduction at threshold in Ref.\,\cite{Ch}.
The same flaw is present also in the earlier study of the pion electroproduction amplitude \cite{BNR}.

Low energy theorems for the amplitude $M^{nj}_\lambda(q,k)$ were derived \cite{ADa},\cite{AFFR},\cite{ITr},\cite{VZ}
using the current conservation and PCAC. These theorems 
have found copious applications (see Refs.\,\cite{H1}--\cite{BKM2}
and references therein). 
%
The relevance of the nucleon weak axial form factor to the pion electroproduction off the nucleon has been 
revisited recently by Haberzettl \cite{H1}. The pion production amplitude \cite{H3},\cite{H4}
was rederived \cite{H1} using an analogue of Eq.\,(\ref{spMnjl}) valid beyond the soft pion limit. 
Neglecting possible Schwinger terms, the amplitude is
\be
M^{nj}_\lambda(q,k)=\frac{q^2+m^2_\pi}{f_\pi\,m^2_\pi}\left[\,i q_\mu\,
\left<p\,'|\int d^4y\, e^{-iqy}\,T\left(\hat{J}^n_{5\mu}(y)\hat{J}^j_\lambda(0)\right)|p\right>
+\vepjnm\left<p\,'|\hat{J}^m_{5\lambda}(0)|p\right>\,\right].   \label{pMnjl}
\ee
The current--current amplitudes were  constructed \cite{H1} by inserting photon lines in all possible 
places in the nucleon weak axial current graph of Fig.\,1.
After calculating the divergence of the current--current amplitudes, the author \cite{H1} obtained
Eq.\,(19), which differs from our Eq.\,(\ref{pMnjl}) by reversing the left- and right hand sides 
and by an additional term, $\bar{u}(p')\,{\cal W}_\lambda\, u(p)$,
accompanying the pion production amplitude in the right hand side of Eq.\,(19). According to the
discussion \cite{H1} after Eq.\,(20), in order for Eq.\,(19) to be consistent
with Eq.\,(\ref{pMnjl}), this term should vanish, 
\be
\bar{u}(p')\,{\cal W}_\lambda\, u(p)\,=\,0\,.  \label{W0}
\ee
The derivation of Eq.(19) inspired Haberzettl to argue that the entire $F_A$
dependence of its left hand side is retained solely in ${\cal W}_\lambda$ on the right hand side, 
since the pion production amplitude $M^{nj}_\lambda(q,k)$ itself does not depend on $F_A$.
Then the conclusion followed \cite{H1} that the pion electroproduction processes at threshold
cannot be used to extract any information regarding the nucleon weak axial form factor.

This claim has been 
criticised \cite{G},\cite{BKM3}, but the critics have been promptly refuted \cite{H2}. 
Actually, the only merit of the criticism made in \cite{G} is that it attracted attention to the problem.
Otherwise, the argument that the contribution to the first term on the 
right hand side of Eq.\,(\ref{spMnjl}) in the soft pion limit is only from the diagrams \ref{figCCA}a and \ref{figCCA}b, is incorrect.
The point is that if one of the currents is axial, then it can couple to the pion or $a_1$ meson lines which emerge
from the vertex (vector current---$B$--boson), and also  
by a direct contact to this vertex. The processes due to the pion emission
contribute to the $B$--pole current--current amplitude by a portion 
\be
\Delta T^{\pi,\,nj}_{\mu \lambda}(B)\,=\,i f_\pi q_\mu \Delta^\pi_F(q^2)\,\Delta M^{nj}_\lambda(B)   \,,\label{dpMnj}
\ee
while the divergence of the current--current amplitudes due to the other two processes yields
\be
iq_\mu \Delta T^{c,\,nj}_{\mu \lambda}(B)\,=\,f_\pi\,\Delta M^{nj}_\lambda(B)\,+\,{\cal C}^{nj}_\lambda(B)\,,  \label{dcMnj}
\ee
where $\Delta M^{nj}_\lambda(B)$ is the associated pion production amplitude and a term ${\cal C}^{nj}_\lambda(B)$
appears, since the chiral group is non--Abelian.
Then the sum of the two contributions to the divergence of the $B$--pole current--current amplitude in (\ref{spMnjl}) is
\be
iq_\mu [\,\Delta T^{\pi,\,nj}_{\mu \lambda}(B)\,+\,\Delta T^{c,\,nj}_{\mu \lambda}(B)\,]\,=
\,f_\pi\,\Delta M^{nj}_\lambda(B)\,+\,{\cal C}^{nj}_\lambda(B)\,,  \label{dtMnj}
\ee
and a new contribution to the pion production amplitude by the vector--isovector current appears
even in the soft pion limit.
As we have already discussed above for the case $B=\pi$ and as we shall see later also for $B=a_1$,
this picture is valid, indeed.

In our opinion, Guichon's argument against the applied formalism is also irrelevant. It is true that the
formalism developed in \cite{H1} for the treatment of the interaction of the axial and vector currents
is not standard, but it is formally correct. 

After making our own study of the problem reported here and comparing it with the earlier results \cite{H1}, 
we came to the conclusion  that the problem is not in the formalism, but is rather with a misinterpretation 
of the results by Haberzettl \cite{H1}.
  
We shall first study the structure of  the pion electroproduction amplitude derived from chiral Lagrangians
based on the local chiral symmetry $SU(2)\times SU(2)$ \cite{STK},\cite{IT1},\cite{T},\cite{STG}. 
These models have recently been used \cite{KT} to construct the one--boson axial  exchange currents for the
Bethe--Salpeter equation and the transition amplitude for the radiative muon capture by proton \cite{STK}.
For the chiral Lagrangians \cite{STK},\cite{STG},  
we shall derive the pion electroproduction amplitude in the same way as it was done in
\cite{H1}: we shall construct the current--current amplitudes and then we shall calculate the divergence. It will become clear that 
this way of constructing the pion production amplitude cannot change its content. Once the amplitude is built  
independently within the same concept (which is the case of the method developed in \cite{H1},\cite{H3},\cite{H4} and
in our approach, too), then to calculate it using the divergence of the current--current amplitudes
is a mere exercise in calculating the diagrams, since one should get the identity.
Our chiral models provide the contact term of the pion production amplitude 
with the nucleon weak axial form factor $F_A(k^2)$ in the soft pion limit, as it should be, since the chiral invariance restricts them
in such a way that they reproduce in the tree approximation predictions of the current algebra and PCAC.
On the other hand, the method of Refs.\,\cite{H1},\cite{H3} and \cite{H4}, avoiding chiral invariance and
based on the gauge invariance only, can produce  in the pion electroproduction amplitude the contact term retaining
in the soft pion limit only the form factor $F^V_1(k^2)$ .

Our pion production amplitude is valid for both soft and hard pions. 
We use first a minimal Lagrangian \cite{IT1}, built  in terms of the Yang--Mills gauge fields, 
for the direct construction of the pion electroproduction amplitude.
For the Lagrangian \cite{STG} reflecting the hidden local $SU(2)\times SU(2)$ symmetry \cite{BKY},\cite{M}, 
we first construct the current--current amplitudes
and then we calculate the divergence. We show that besides the amplitudes of Fig.\,\ref{figCCA}a and \ref{figCCA}b
both exchanges, $\pi$ and $a_1$, contribute to the amplitude of Fig.\,\ref{figCCA}. As a result, we 
obtain the pion production amplitude
of our model minus the last term on the right hand side of Eq.\,(\ref{spMnjl}), as it should be, if the calculations are
consistent. On the other hand, the divergence of the pion pole current--current amplitude provides the standard
pion pole production amplitude and the divergence of the $a_1$ meson pole current--current amplitude produces 
the new contact term containing again the nucleon weak axial form factor $F_A(k^2)$ and other terms
of the order at least ${\cal O}(q)$ in the soft pion limit.

In sect.\,\ref{CH1}, we introduce the necessary formalism, in sect.\,\ref{CH2} we construct the pion production amplitude. Our
results and conclusions are summarized in sect.\,\ref{CH3}.

\section{Formalism}
\label{CH1}

Direct construction of the amplitude $M^{nj}_\lambda(q,k)$ will be carried out using the minimal Lagrangian \cite{IT1},\cite{T}.
Here we write only  the necessary vertices

\begin{eqnarray}
\Delta{\cal L}^{YM}_{N \pi \rho\,a_1} & = & -{\bar N}\gamma_\mu
\partial_\mu N -M{\bar N}N-i\frac{g}{2M}{\bar N} \not{\partial}
(\vec{\Pi}\,\cdot\,\vec{\tau}) \gamma_5 N
+ ig_\rho \frac{g_A}{2f_\pi}{\bar N}\gamma_\mu \gamma_5
(\vec{\tau}\,\cdot\,\vec{\rho}_\mu \times \vec{\Pi})N \nonumber  \\
& & -ig_A g_\rho {\bar N}\gamma_\mu \gamma_5 (\vec{\tau}\,\cdot\,
\vec a_\mu)N  -i\frac{g_\rho}{2}{\bar N}(\gamma_\mu \vec{\rho}_\mu
-i\frac{\kappa_V}{4M}\sigma_{\mu \nu}\vec{\rho}_{\mu \nu})
\,\cdot\,\vec{\tau}N  \nonumber \\
& & + g_\rho \vec{\rho}_\mu \cdot \vec{\Pi} \times \partial_\mu \vec{\Pi} 
- \frac{1}{f_\pi} (\vec{\rho}_{\mu \nu} \cdot \vec{a}_\mu \times 
\partial_\nu \vec \Pi + \frac{1}{4} \vec{\rho}_{\mu \nu}\cdot
\vec{\Pi} \times \vec{a}_{\mu \nu})\,, \label{dYML}
\end{eqnarray}
where
\begin{equation}
\vec{\rho}_{\mu \nu} = \partial_\mu \vec{\rho}_\nu - \partial_\nu \vec{\rho}_\mu\,.
\end{equation}

From the associated one--body currents we present only the vector--isovector current
\be
J^j_{YM,\lambda} = -\frac{m^2_\rho}{g_\rho}\,\rho^j_\lambda
\,.  \label{JjYM} 
\ee

We use the Lagrangian model \cite{STK},\cite{STG} reflecting the hidden local symmetry
to derive the amplitude  $M^{nj}_\lambda(q,k)$ employing Eq.\,(\ref{pMnjl}). 
It reads in the needed approximation
\begin{eqnarray}
{\cal L}^{HLS}_{N \pi \rho\,a_1} & = & -{\bar N}\gamma_\mu
\partial_\mu N -M{\bar N}N+ig{\bar N} \gamma_5
(\vec{\Pi}\,\cdot\,\vec{\tau})N
- ig_\rho \frac{g_A}{2f_\pi}{\bar N}\gamma_\mu \gamma_5
(\vec{\tau}\,\cdot\,\vec{\rho}_\mu \times \vec{\Pi})N \nonumber  \\
& & -i\frac{g_\rho g^2_A}{f_\pi} {\bar N} \gamma_\mu(\vec{\tau}\,\cdot\,
 {\vec a}_\mu \times \vec{\Pi})N
-ig_A g_\rho {\bar N}\gamma_\mu \gamma_5 (\vec{\tau}\,\cdot\,
\vec a_\mu)N  \nonumber \\
& & -i\frac{g_\rho}{2}{\bar N}(\gamma_\mu \vec{\rho}_\mu
-i\frac{\kappa_V}{4M}\sigma_{\mu \nu}\vec{\rho}_{\mu \nu})
\,\cdot\,\vec{\tau}N 
+i\frac{g_\rho g_A}{4f_\pi}\frac{\kappa_V}{2M}\,{\bar
N}\gamma_5 \sigma_{\mu \nu} (\vec{\Pi}\cdot \vec{\rho}_{\mu \nu})N
\nonumber \\
& & + g_\rho \vec{\rho}_\mu \cdot \vec{\Pi} \times \partial_\mu \vec{\Pi} 
-g_\rho \partial_\nu \vec{\rho}_\mu \cdot \vec{\rho}_\mu \times 
\vec{\rho}_\nu + g_\rho (\vec{\rho}_\mu \times \vec{a}_\nu -
\vec{\rho}_\nu \times \vec{a}_\mu) \cdot \partial_\mu \vec{a}_\nu
\nonumber \\
& & + \frac{1}{f_\pi} (\vec{\rho}_{\mu \nu} \cdot \vec{a}_\mu \times 
\partial_\nu \vec \Pi + \frac{1}{2} \vec{\rho}_\mu \cdot
\partial_\nu \vec{\Pi} \times \vec{a}_{\mu \nu})\,. \label{HLSL}
\end{eqnarray}

The associated one--body currents are 
\begin{eqnarray}
J^j_{HLS\,,\,\lambda} & = & -\frac{m^2_\rho}{g_\rho}\, 
\rho^j_\lambda-2f_\pi g_\rho \, \left({\vec a}_\lambda
\times \vec{\Pi}\right)^j+{\cal O}(|\vec{\Pi}|^2)\,,  \label{JjHLS}  \\
J^n_{HLS\,,5\mu} & = &
-\frac{m^2_\rho}{g_\rho}\,a^n_\mu
+f_\pi \partial_\mu \Pi^n-2f_\pi g_\rho\,
\left(\vec{\rho}_\mu \times \vec{\Pi}\right)^n \nonumber \\
& & +\frac{1}{g_\rho}\,\left
[(\frac{1}{2f_\pi}\, \partial_\nu \vec{\Pi}
-g_\rho \vec{a}_\nu + e\vec{\cal A}_\nu) \times
\vec{\rho}_{\mu \nu}\right]^n \,. \label{J5nHLS}  
\end{eqnarray}

\section{Pion electroproduction amplitude}
\label{CH2}

This section is devoted to the construction of the pion production amplitude starting from the chiral
Lagrangians and currents of the previous section and using the standard Feynman graph technique.

\subsection{Pion electroproduction amplitude from the minimal chiral Lagrangian}
\label{CH2.1}

Generally, the amplitude is graphically presented as in Fig.\,\ref{figSPPA}. The Born terms, Fig.\,\ref{figSPPA}a and 
\ref{figSPPA}b, are of the same form as the first two terms on the right hand side of Eq.\,(\ref{AME}), only the 
current ${\hat J}^j_\lambda(k)$ is given \cite{STK} by the vector dominance model,
\be
\hat J^j_\lambda(k)  =  i m^2_\rho \Delta^\rho_{\lambda \zeta}
(k)(\gamma_\zeta - \frac{\kappa_V}{2M} \sigma_{\zeta \delta}
k_\delta)\,\frac{\tau^j}{2}\,. \label{JjV}
\ee
Then the Born term is,
\be
M^{nj}_{B,\,\lambda}\,=\,\frac{g}{2M}\,\bar{u}(p\,')\left[\not q \tau^n S_F(P) \hat{J}^j_\lambda(k)
\,+\,\hat{J}^j_\lambda(k) S_F(Q) \not q \tau^n\right]u(p)\,.  \label{YMBT}
\ee
Also the pion pole production term, Fig.\,\ref{figSPPA}c, is of the standard form with the pion form factor provided 
again by the vector dominance model
\be
M^{nj}_{pp,\,\lambda}\,=\,m^2_\rho\,\Delta^\rho_{\lambda\zeta}(k)\,(\,q_{1\zeta}\,-\,q_\zeta\,)\,\Delta^\pi_F(q^2_1)
\vepjnm\,g\Gamma^m_5(p',p)\,,  \label{YMPPT}
\ee
where
\be
\Gamma^m_5(p',p)\,  =\,  \bar u(p')\, \g5 \tau^m\, u(p)\,.
\label{G5m}
\ee

\begin{figure}[h!]
\centerline{
\epsfig{file=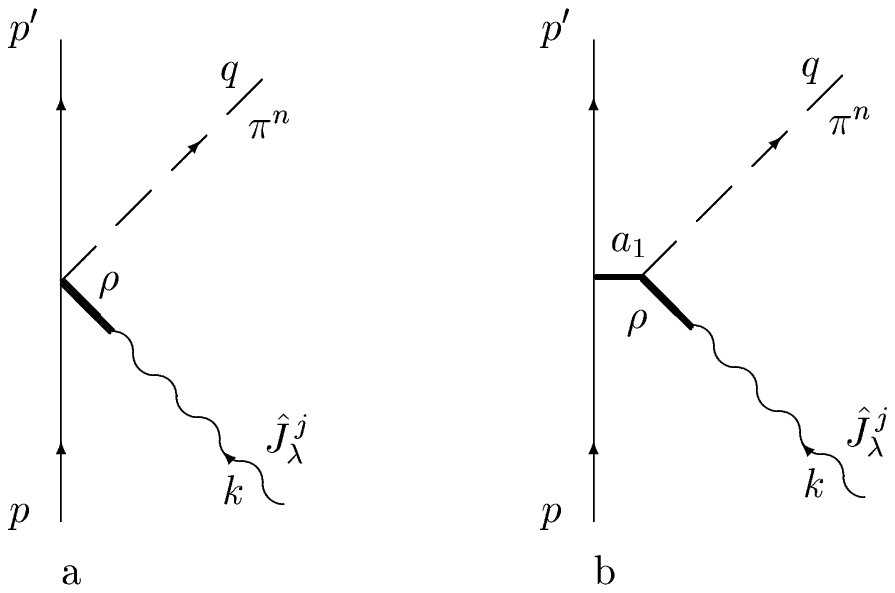}
}
\vskip 0.4cm
\caption{
The contact terms arising from the chiral Lagrangian Eq.\,(\ref{dYML}).
Graph a corresponds to the fourth term, graph b is constructed from the last two terms.
}
\label{figYMCT}
\end{figure}

However, the contact term, Fig.\,\ref{figSPPA}d, deserves more attention. Visually, it is
given by two graphs of Fig.\,\ref{figYMCT}a and \ref{figYMCT}b.
The amplitude corresponding to the graph Fig.\,\ref{figYMCT}a is
\be
M^{nj}_{c_1,\,\lambda}\,=\,i\frac{g_A}{2f_\pi}\,m^2_\rho\Delta^\rho_{\lambda \zeta}(k)\,\vepjnm\,
\Gamma^m_{5\zeta}(p',p)\,
\rightarrow\,i\frac{g_A}{2f_\pi}\,F^V_1(k^2)\vepjnm\,\Gamma^m_{5\lambda}(p',p)\,, \label{SCT}
\ee
where
\be
\Gamma^m_{5\lambda}(p',p)\,  =\,  \bar u(p')\, \gamma_\lambda \g5 \tau^m\, u(p)\,.
\label{G5lm}
\ee

This is the Kroll--Ruderman term in the pion production amplitude, which can be obtained
by the minimal substitution \cite{Fr}
\begin{displaymath}
\partial_\lambda\,\longrightarrow\,\partial_\lambda\,\pm\,ie\,F^V_1\,A_\lambda\,,
\end{displaymath}
in the pseudovector $\pi NN$ interaction [the 3rd term on the right hand side of Eq.\,(\ref{dYML})].
Let us now present the contact term, Fig.\,\ref{figYMCT}b, arising from the last two vertices of Eq.\,(\ref{dYML})
\be
M^{nj}_{c_2,\,\lambda}\,=\,i\,\frac{g_A}{f_\pi}\,\vepjnm\,m^2_\rho[\,k_\nu\,\Delta^\rho_{\lambda \eta}(k)
\,-\,k_\eta\,\Delta^\rho_{\lambda \nu}(k)\,]\,(\,q_\nu\,+\,\frac{1}{2}q_{1\,\nu}\,)
\,\Delta^{a_1}_{\eta \zeta}(q_1)\,\Gamma^m_{5\zeta}(p',p)\,.   \label{ACT}
\ee
This amplitude is transverse by itself.
Taking into account that $q_1=k-q$, we can write
\be
M^{nj}_{c_2,\,\lambda}\,=\,i\,\frac{g_A}{2f_\pi}\,m^2_\rho\,\Delta^\rho_F(k^2)
\Delta^{a_1}_F(k^2)\,k^2\,\vepjnm\,\Gamma^m_{5\lambda}(p',p)\,+\,\Delta M^{nj}_{c_2\,,\lambda}\,,   \label{ACTf}
\ee
where in the soft pion limit
\be
\Delta M^{nj}_{c_2,\,\lambda}\,=\,{\cal O}(q,k^2)\,.  \label{dACT}
\ee
Summing up the amplitudes given in Eq.\,(\ref{SCT}) and (\ref{ACTf}), we get the total contact term $M^{nj}_{c\,,\lambda}$
of this model in the soft pion limit
\be
M^{nj}_{c,\,\lambda}\,=\,i\frac{g_A}{2f_\pi}\,F_A(k^2)\vepjnm\,\Gamma^m_{5\lambda}(p',p)
\,+\,{\cal O}(q,k^2)\,.   \label{SACT}
\ee
In deriving Eqs.\,(\ref{SCT})--(\ref{SACT}), we used the Weinberg relation $m^2_{a_1}=2m^2_\rho$ \cite{W} and 
the vector dominance model form of the form factors $F^V_1(k^2)$ and $F_A(k^2)$ which is incorporated in
the model Lagrangian Eq.\,(\ref{dYML}).

The contact term, Eq.\,(\ref{SACT}), coincides in form with the last term 
of the soft pion amplitude Eq.\,(\ref{sppa}). Let us stress that the nucleon weak axial form factor $F_A(k^2)$ appears
in the contact term in order to satisfy the local chiral invariance.

The gauge condition for the pion electroproduction amplitude of this model, given by 
\mbox{Eqs.\,(\ref{YMBT})--(\ref{ACT})}, is in agreement with the 
Ward--Takahashi relation \cite{ChS}
\be
k_\lambda\left[\,M^{nj}_{B,\,\lambda}+M^{nj}_{pp,\,\lambda}+M^{nj}_{c_1,\,\lambda}+M^{nj}_{c_2,\,\lambda}\,\right]\,=\,
-\frac{m^2_\pi+q^2}{m^2_\pi+q^2_1}\,\vepjnm\,g\Gamma^m_{5}(p',p)\,.   \label{GG}
\ee

\subsection{Pion electroproduction amplitude from the hidden local symmetry Lagrangian}
\label{CH2.2}

In this section, the pion electroproduction amplitude will be obtained via Eq.\,(\ref{pMnjl}), and for constructing
the current--current amplitudes we use the chiral Lagrangian, Eq.\,(\ref{HLSL}), and the associated currents, 
Eq.\,(\ref{JjHLS}) and (\ref{J5nHLS}). This formalism has been employed in \cite{STK} for the derivation of the 
current--current amplitudes for the radiative muon capture by protons. 
Those amplitudes and the ones we need to calculate here are related by time reversal. 
The minimal amplitudes, direct analogues of the current--current amplitudes
of Fig.\,3 of Ref.\,\cite{H1}, were derived for the radiative muon capture \cite{RoT} almost
40 years ago and they were used directly in \cite{F},\cite{BF1} for calculating the transition rate of the reaction
\be 
\mu\,+\,p\,\longrightarrow\,n\,+\,\nu_\mu\,+\gamma\,.   \label{RMC}
\ee
The case of the pion electroproduction is simpler, since only 
the vector-- and the axial--isovector currents contribute.

\subsubsection{Nucleon Born terms}
\label{CH2.2.1}

\begin{figure}[h!]
\centerline{
\epsfig{file=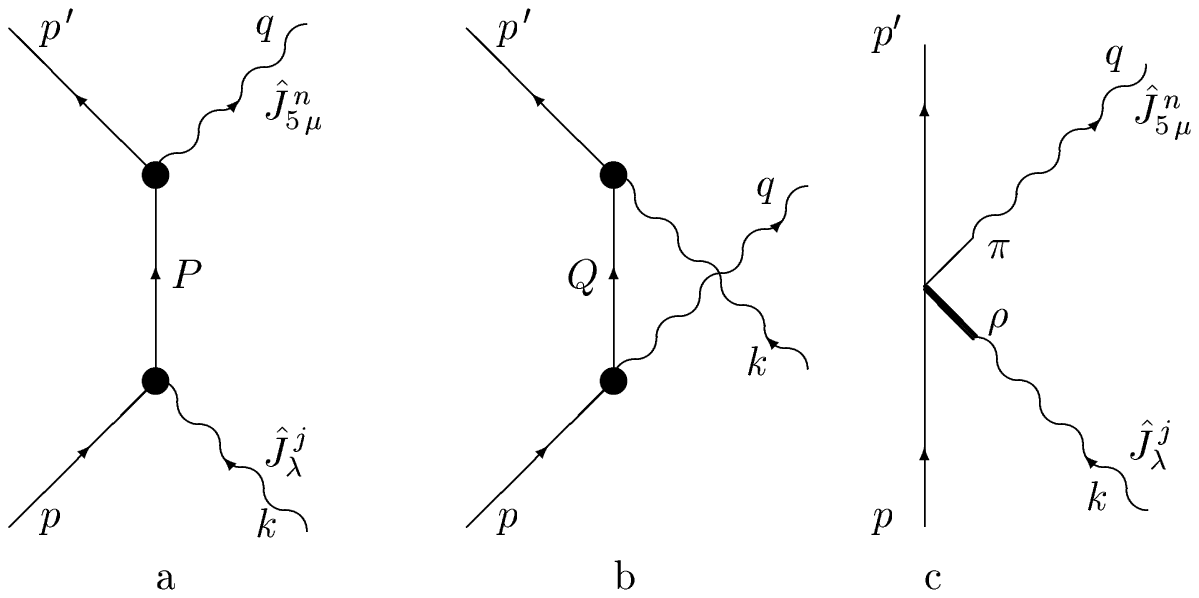}
}
\vskip 0.4cm
\caption{
The nucleon Born current--current terms given by the Lagrangian Eq.\,(\ref{HLSL}) and the currents Eq.\,(\ref{JjHLS})
and (\ref{J5nHLS}).
}
\label{figBCCT}
\end{figure}

Let us start with the nucleon Born terms. According to our Lagrangian Eq.\,(\ref{HLSL}), we use the chiral 
model with the pseudoscalar $\pi NN$ coupling and  besides the graphs Fig.\,\ref{figBCCT}a and \ref{figBCCT}b,
the contact term Fig.\,\ref{figBCCT}c should be calculated.
There are three nucleon Born current--current amplitudes
\bea 
T^{B,\,nj}_{\mu \lambda}(1) & = & -{\bar u}(p')[\,\hat J^n_{5\,\mu}(q) 
S_F(Q)\hat J^{j}_\lambda(k)\,+\,\hat J^{j}_\lambda(k) S_F(P) \hat 
J^n_{5\,\mu}(q)\,]u(p)  \label{TBnj1} \\ 
T^{B,\,nj}_{\mu \lambda}(2) & = & 
-\frac{g_A}{2}  q_\mu \Delta^\pi_F(q^2) m^2_\rho \Delta^\rho_{\lambda \zeta}(k) 
\vepjnm \,\Gamma^m_{5\zeta}(p',p)\, \nonumber \\ &\equiv & 
\, i f_\pi q_\mu \Delta^\pi_F(q^2)\, M^{nj}_{B,\,\lambda}(2)\,, \label{TBnj2} \\ 
T^{B,\,nj}_{\mu \lambda}(3) & = & i\frac{g_A}{2}  q_\mu \Delta^\pi_F(q^2)
\frac{\kappa_V}{2M} m^2_\rho 
\Delta^\rho_{\lambda \eta}(k) k_\zeta \delta_{j\,n} \bar 
u(p')\gamma_5\,\sigma_{\zeta \eta} u(p) \nonumber \\ & \equiv & \, i f_\pi 
q_\mu \Delta^\pi_F(q^2)\,M^{nj}_{B,\,\lambda}(3)\,. \label{TBnj3} 
\eea

The contribution of the induced pseudoscalar part of the axial current 
$\hat J^n_{5\,\mu}(q)$ to the amplitude $T^{B,\,nj}_{\mu \lambda}(1)$ is
\bea
T^{B,\,nj}_{\mu \lambda}(\pi,1)\,&=&\,f_\pi\,{\bar u}(p')[\,g\g5\,\tau^n 
S_F(Q)\hat J^{j}_\lambda(k)\,+\,\hat J^{j}_\lambda(k) S_F(P) g\g5\,\tau^n\,]u(p)\,
\nonumber \\ & \equiv & \,i f_\pi 
q_\mu \Delta^\pi_F(q^2)\,M^{nj}_{B,\,\lambda}(1)\,. \label{TBnjpi1} 
\eea

Calculation of the divergence of the amplitude $T^{B,\,nj}_{\mu \lambda}(1)$, Eq.\,(\ref{TBnj1}), yields
\be
iq_\mu T^{B,\,nj}_{\mu \lambda}(1)\,=\,f_\pi m^2_\pi\Delta^\pi_F(q^2)\,M^{nj}_{B,\,\lambda}(1)
\,+\,f_\pi\sum_{i=2}^3 M^{nj}_{B,\,\lambda}(i)\,.  \label{dTBnj1}
\ee
Then it follows from Eqs.\,(\ref{TBnj2}), (\ref{TBnj3}) and (\ref{dTBnj1}) that
\be
iq_\mu\, \sum_{i=1}^3 T^{B,\,nj}_{\mu \lambda}(i)\,=\,
f_\pi m^2_\pi\,\Delta^\pi_F(q^2)\,\sum_{i=1}^3 M^{nj}_{B,\,\lambda}(i)\,
\equiv\,f_\pi m^2_\pi\,\Delta^\pi_F(q^2)\,M^{nj}_{B,\,\lambda}\,,    \label{dTBnj}
\ee
where $M^{nj}_{B,\,\lambda}$ is the nucleon Born pion electroproduction amplitude for the pseudovector 
$\pi NN$ coupling, Eq.\,(\ref{YMBT}). The amplitudes $M^{nj}_{B,\,\lambda}(2)$ and $M^{nj}_{B,\,\lambda}(3)$ 
ensure the PCAC constraint.

\subsubsection{Pion pole terms}
\label{CH2.2.2}
\begin{figure}[h!]
\centerline{
\epsfig{file=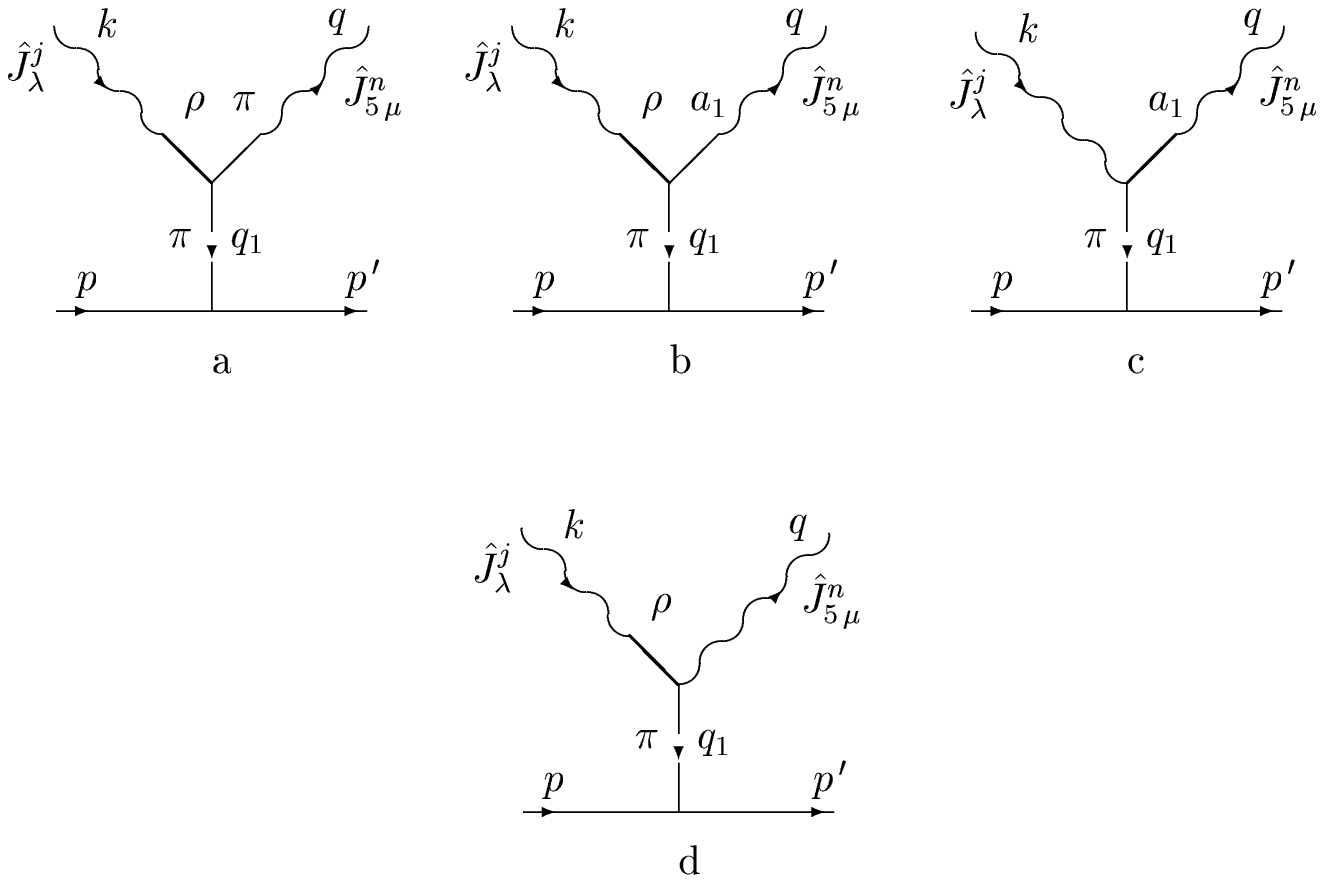}
}
\vskip 0.4cm
\caption{
The pion pole current--current amplitudes given by the Lagrangian Eq.\,(\ref{HLSL}) and the currents Eq.\,(\ref{JjHLS})
and (\ref{J5nHLS}).
}
\label{figPPT}
\end{figure}
The current--current amplitudes due to the pion exchange in Fig.\,\ref{figCCA}c are presented 
in Figs.\,\ref{figPPT}a--\ref{figPPT}d.
They are
\bea
T^{pp,\,nj}_{\mu \lambda}(a)\,&=&\,if_\pi q_\mu \Delta^\pi_F(q^2) m^2_\rho
\Delta^\rho_{\eta \lambda}(k)\,(q_{1 \eta}-q_\eta)
\Delta^\pi_F(q^2_1)\,\vepjnm\,g\Gamma^m_5(p',p) \nonumber  \\
& \equiv & \,if_\pi q_\mu \Delta^\pi_F(q^2)
M^{nj}_{pp,\,\lambda}\,, \label{TPPnja} \\
T^{pp,\,nj}_{\mu \lambda}(b)\,&=&\,  2i f_\pi m^2_\rho \Delta^{a_1}_{\mu
\eta}(q)\Delta^\rho_F(k^2)(\,q_1 \cdot k\,\delta_{\eta \lambda}\,-\,
k_\eta q_{1 \lambda}\,)\,\Delta^\pi_F(q^2_1)\,\vepjnm\,g\Gamma^m_5(p',p) \nonumber\\
&& \,+\,if_\pi  m^2_\rho \Delta^{a_1}_F(q^2)
\Delta^\rho_{\eta \lambda}(k)(\,q \cdot q_1\,\delta_{\eta \mu}\,-\,
q_\eta q_{1 \mu} \,)\,\Delta^\pi_F(q^2_1)\,\vepjnm\,g\Gamma^m_5(p',p)\,,
\label{TPPnjb} \\
T^{pp,\,nj}_{\mu \lambda}(c)\,&=&\, - 2if_\pi  m^2_\rho \Delta^{a_1}_{\mu
\lambda}(q)\,\Delta^\pi_F(q^2_1))\,\vepjnm\,g\Gamma^m_5(p',p)\,,
\label{TPPnjc} \\
T^{pp,\,nj}_{\mu \lambda}(d)\,&=&\,2i f_\pi m^2_\rho \Delta^\rho_{\mu
\lambda}(k)\,\Delta^\pi_F(q^2_1)\,\vepjnm\,g\Gamma^m_5(p',p)\,\nonumber \\
&& \,+i f_\pi \Delta^\rho_F(k^2)(\,k_\mu q_{1 \lambda}\,-\,
q_1 \cdot k\,\delta_{\mu \lambda}\,)\,\Delta^\pi_F(q^2_1)\,\vepjnm\,g\Gamma^m_5(p',p)\,.
\label{TPPnjd} 
\eea
The process of Fig.\,\ref{figPPT}a is the only one, where the axial current is attached to the pion emitted from the
vertex (vector current---$B$--boson). The associated pion production 
amplitude $M^{nj}_{pp,\,\lambda}$ is given by the same graph without the axial current wavy line.

Let us first calculate the divergence of the amplitudes Eqs.\,(\ref{TPPnjb})--(\ref{TPPnjd}). It is easy to find that 
the divergence of the first part of the amplitude Eq.\,(\ref{TPPnjb}) and of the second part of the amplitude
Eq.\,(\ref{TPPnjd}) cancel. The rest provides
\be
iq_\mu\,\sum_{x=b,c,d} T^{pp,\,nj}_{\mu \lambda}(x)\,=\, 
f_\pi M^{nj}_{pp,\,\lambda}\,-\,f_\pi q_{1\,\lambda}\Delta^\pi_F(q^2_1)\,\vepjnm\,g\Gamma^m_5(p',p) 
\,, \label{dTPPnjbcd}
\ee
where $M^{nj}_{pp,\,\lambda}$ is the pion pole production amplitude defined in Eq.\,(\ref{TPPnja}), and it coincides 
with the amplitude Eq.\,(\ref{YMPPT}) of the section \ref{CH2.1}. Let us note that Eq.\,(\ref{dTPPnjbcd}) is 
a particular form of the general result given in Eq.(\ref{dcMnj}).
 
Then the divergence of the pion pole current--current amplitudes Eqs.\,(\ref{TPPnja})--(\ref{TPPnjd}) is
\be
iq_\mu\,\sum_{x=a,b,c,d} T^{pp,\,nj}_{\mu \lambda}(x)\,=\, 
f_\pi m^2_\pi\Delta^\pi_F(q^2)\,M^{nj}_{pp,\,\lambda}\,-\,f_\pi q_{1\,\lambda}
\Delta^\pi_F(q^2_1)\,\vepjnm\,g\Gamma^m_5(p',p) \,. \label{dTPPnjabcd}
\ee
Using this result and Eq.\,(\ref{JAs}) for the matrix element
of the nucleon weak axial current, we immediately see from Eq.\,(\ref{pMnjl}) that 
the induced pseudoscalar disappears from the pion electroproduction amplitude and the pion pole production
amplitude $M^{nj}_{pp,\,\lambda}$, Eq.\,(\ref{YMPPT}), appears instead, as already discussed above. 
Evidently, this statement is independent of the pion being soft or hard.

\subsubsection{$a_1$ meson pole terms}
\label{CH2.2.3}
\begin{figure}[h!]
\centerline{
\epsfig{file=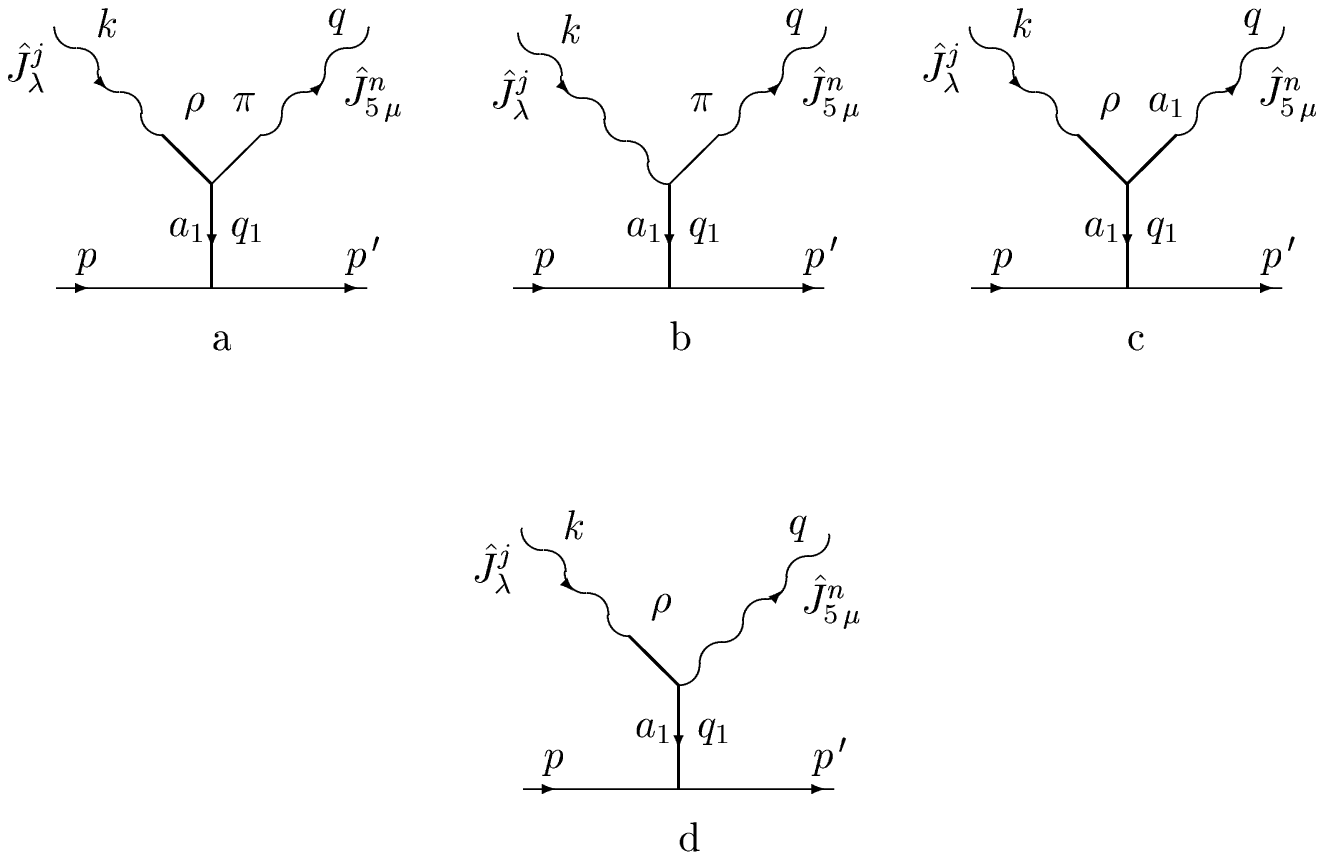}
}
\vskip 0.4cm
\caption{
The $a_1$ meson pole current--current amplitudes given by the Lagrangian Eq.\,(\ref{HLSL}) and the currents Eq.\,(\ref{JjHLS})
and (\ref{J5nHLS}).
}
\label{figA1PT}
\end{figure}
The contributions to the current--current amplitude from the $a_1$ meson pole term, Fig.\,\ref{figCCA}c,
are given in Figs.\,\ref{figA1PT}a--\ref{figA1PT}d. They have the form
\bea
T^{a_1p,\,nj}_{\mu \lambda}(a)\,&=&\,g_A m^2_\rho\,q_\mu\,\Delta^\pi_F(q^2)\,\left\{\,\Delta^{a_1}_{\nu \zeta}(q_1)\,
\Delta^\rho_F(k^2)\,[\,(k\cdot q)\delta_{\nu \lambda}\,-\,k_\nu q_\lambda\,]\right. \nonumber  \\
&&  \left.\, +\,\frac{1}{2}\Delta^\rho_{\nu \lambda}(k)\,\Delta^{a_1}_F(q^2_1)\,[\,(q\cdot q_1)\delta_{\zeta \nu}
\,-\,q_\zeta q_{1\eta}\,]\,\right\}\,\vepjnm\, \Gamma^m_{5\zeta}(p',p) \nonumber  \\
&\equiv& \,i f_\pi\, q_\mu\, \Delta^\pi_F(q^2)\,M^{nj}_{a_1p,\,\lambda}(a)\,, \label{TA1Pnja} \\
T^{a_1p,\,nj}_{\mu \lambda}(b)\,&=&\,-g_A m^2_\rho\,q_\mu\,\Delta^\pi_F(q^2)\,\Delta^{a_1}_{\lambda \zeta}(q_1)
\,\vepjnm\, \Gamma^m_{5\zeta}(p',p) \nonumber  \\
&\equiv& \,i f_\pi\, q_\mu\, \Delta^\pi_F(q^2)\,M^{nj}_{a_1p,\,\lambda}(b)\,, \label{TA1Pnjb} \\
T^{a_1p,\,nj}_{\mu \lambda}(c)\,&=&\,g_A m^4_\rho\,\Delta^\rho_{ \lambda \eta}(k)
\,[\,(q_{1 \eta}-q_\eta)\Delta^{a_1}_{\mu \nu}(q)\,
\Delta^{a_1}_{\nu \zeta}(q_1)
+q_\nu \Delta^{a_1}_{\mu \eta}(q)\, \Delta^{a_1}_{\nu
\zeta}(q_1) \nonumber \\
& &\, -\, q_{1 \nu} \Delta^{a_1}_{\mu \nu}(q)\,
\Delta^{a_1}_{\zeta \eta}(q_1)\,]\,
\vepjnm\, \Gamma^m_{5\zeta} (p',p)\,, \label{TA1Pnjc}  \\
T^{a_1p,\,nj}_{\mu \lambda}(d)\,&=&\,g_A m^2_\rho\,\Delta^\rho_F(k^2)\,
[\,k_\eta\delta_{\mu \lambda}\,-\,k_\mu\delta_{\lambda \eta}\,]
\Delta^{a_1}_{\eta \zeta}(q_1)\,\vepjnm\, \Gamma^m_{5\zeta} (p',p)\,. \label{TA1Pnjd}
\eea
The two processes, Fig.\,\ref{figA1PT}a and \ref{figA1PT}b, contribute to the pion production
amplitude providing the terms $M^{nj}_{a_1p,\,\lambda}(a)$ and $M^{nj}_{a_1p,\,\lambda}(b)$,
which are given in Eq.\,(\ref{TA1Pnja}) and (\ref{TA1Pnjb}), respectively.

Let us calculate the divergence of the sum of the amplitudes Eq.\,(\ref{TA1Pnjc}) and (\ref{TA1Pnjd}).
The result is
\be
iq_\mu\,\sum_{x=c,d}\,T^{a_1p,\,nj}_{\mu \lambda}(x)\,=\,f_\pi\,\sum_{x=a,b}\,M^{nj}_{a_1p,\,\lambda}(x)\,
-\,ig_A m^2_\rho\,\Delta^{a_1}_{\lambda \zeta}(q_1)\,\vepjnm\, \Gamma^m_{5\zeta} (p',p)\,. \label{dTA1Pcd}
\ee
Then it is clear that the divergence of the sum of all the $a_1$ meson pole current--current
amplitudes, given by Eqs.\,(\ref{TA1Pnja})--(\ref{TA1Pnjd}), is
\be
iq_\mu\sum_{x=a,b,c,d}\,T^{a_1p,\,nj}_{\mu \lambda}(x)=f_\pi\,m^2_\pi\,\Delta^\pi_F(q^2)
\sum_{x=a,b}\,M^{nj}_{a_1p,\,\lambda}(x)
-ig_A m^2_\rho\,\Delta^{a_1}_{\lambda \zeta}(q_1)\,\vepjnm\, \Gamma^m_{5\zeta} (p',p)\,. \label{dTA1Pabcd}
\ee
Substituting this result into Eq.\,(\ref{pMnjl}) we find that the first term in the matrix element of the
axial current, Eq.\,(\ref{JAs}), is cancelled. However, it does not mean that the dependence of the pion electroproduction
amplitude on the nucleon weak axial form factor is eliminated. This will become clear in the next
section, where we present the resulting pion production amplitude. 

\subsubsection{Resulting pion electroproduction amplitude}
\label{CH2.2.4}

Now we present the pion electroproduction amplitude obtained  in Sects.\,\ref{CH2.2.1}--\ref{CH2.2.3} 
from the Lagrangian model Eq.\,(\ref{HLSL}). For this purpose, we insert the results for the divergence of the
current--current amplitudes, Eqs.\,(\ref{dTBnj}),(\ref{dTPPnjabcd}) and (\ref{dTA1Pabcd}), into Eq.\,(\ref{pMnjl}).
The resulting pion electroproduction amplitude is
\be
M^{nj}_\lambda(q,k)\,=\,M^{nj}_{B,\,\lambda}\,+\,M^{nj}_{pp,\,\lambda}\,+\,\sum_{x=a,b}\,M^{nj}_{a_1p,\,\lambda}(x)\,.
\label{Mnjlt}
\ee
It is clear that the matrix element of the axial current is cancelled.  
The amplitudes $M^{nj}_{B,\,\lambda}$ and $M^{nj}_{pp,\,\lambda}$ are the well--known nucleon Born and pion pole
terms, respectively. So only the contribution of the $a_1$ meson pole to the divergence of  the current--current amplitudes
remains to be considered. Let us write the last two terms on the right hand side of Eq.\,(\ref{Mnjlt}) explicitly. From the definitions
in Eq.\,(\ref{TA1Pnja}) and (\ref{TA1Pnjb}) we have
\bea
M^{nj}_{a_1p,\,\lambda}(a)\,&=&\,i\frac{g_A}{2f_\pi}\,m^2_{a_1}\,\left\{\,\Delta^{a_1}_{\nu \zeta}(q_1)\,
\Delta^\rho_F(k^2)\,[\,k_\nu q_\lambda-(k\cdot q)\delta_{\nu \lambda}\,]\right. \nonumber  \\
&&  \left.\, +\,\frac{1}{2}\Delta^\rho_{\nu \lambda}(k)\,\Delta^{a_1}_F(q^2_1)\,[\,q_\zeta q_{1\eta}-(q\cdot q_1)
\delta_{\zeta \nu}\,]\,\right\}\,\vepjnm\, \Gamma^m_{5\zeta}(p',p)\,, \label{Mnja1pa}  \\
M^{nj}_{a_1p,\,\lambda}(b)\,&=&\,i\frac{g_A}{2f_\pi}\,m^2_{a_1}\,\Delta^{a_1}_{\lambda \zeta}(q_1)
\,\vepjnm\, \Gamma^m_{5\zeta}(p',p)\,. \label{Mnja1pb}
\eea
In the soft pion limit, the amplitude $M^{nj}_{a_1p,\,\lambda}(a)\sim {\cal O}(kq)$, while the amplitude
$M^{nj}_{a_1p,\,\lambda}(b)$ restores the contact term retaining the nucleon weak axial form factor $F_A$.
For its construction, the presence of the contact current [the second term on the right hand side of Eq.\,(\ref{JjHLS})]
is decisive.
So this model is also consistent with predictions of the current algebra and PCAC.

Let us emphasize that the amplitude $M^{nj}_\lambda(q,k)$, Eq.\,(\ref{Mnjlt}), satisfies the gauge condition,
expressed in Eq.\,(\ref{GG}).

Of course, the amplitude $M^{nj}_\lambda(q,k)$ obtained from the Lagrangian Eq.\,(\ref{HLSL}) directly, does coincide with
the one obtained from Eq.\,(\ref{Mnjlt}). However, the construction of this section allows us to see how the pion and $a_1$ 
meson poles contribute to the divergence of the current--current amplitudes.

\section{Discussion and summary}
\label{CH3}

We have studied the structure of the pion electroproduction amplitude obtained from the Lagrangians
incorporating the local chiral symmetry. From the  minimal Lagrangian Eq.\,(\ref{dYML}), constructed
in terms of the Yang--Mills heavy meson fields, and using the current
Eq.\,(\ref{JjYM}), we obtained the pion production amplitude directly. 
Actually, the model provides two contact terms.
One of them is prescribed by the gauge invariance and it contains the form factor $F^V_1$.
Another one is due to the gauge chiral invariance. This term is transverse by itself.
Combining these contact terms results in another contact
term containing the nucleon weak axial form factor and a piece of the order ${\cal O}(qk^2)$
in the soft pion limit.

The Lagrangian Eq.\,(\ref{HLSL}), reflecting the hidden local
symmetry, helped us to construct the current--current amplitudes of Fig.\,\ref{figCCA}. Subsequent calculation
of the divergence
of these amplitudes and the use of Eq.\,(\ref{pMnjl}) led us to observe
that the pion and $a_1$ exchanges in Fig.\,\ref{figCCA}c do contribute non--negligibly even in the soft pion limit.
The contribution proceeds in such a way that the matrix element of the axial current [the second term
on the right hand side of Eq.\,(\ref{pMnjl})] is eliminated and the pion pole and $a_1$ pole pion
production terms appear. Moreover, one of the $a_1$ pole pion production amplitudes is nothing but the Kroll--Ruderman term
containing again the nucleon weak axial form factor $F_A(k^2)$ in the soft pion limit.
It was noted many years ago in \cite{Ad} that this contact term originated from the $a_1$ meson exchange.

It is to be noted that the derivation of the
pion production amplitude via Eq.(\ref{pMnjl}) does not supply any additional dynamical input that is not already
present in the considered model and we get
the pion production amplitude Eq.\,(\ref{Mnjlt}), which is the same as the one obtained from the Lagrangian Eq.\,(\ref{HLSL})
by the direct construction.

It follows from our results that the pion electroproduction amplitude does not contain the induced pseudoscalar part
of the nucleon weak axial current either for soft or hard pions, which is at variance with \cite{G}.
It also means that the measurement of this quantity  \cite{Ch} is a misconception.
 
Let us now compare our results with those of Ref.\,\cite{H1}. In this paper, the formalism of the vector current--axial current
interaction is formulated in such a way that the divergence of the current--current amplitudes, where the 
axial current is attached to the pion line, provides the whole pion production amplitude.
As discussed  \cite{H1} in connection with Eq.\,(19)  for the pion electroproduction amplitude,
Eq.\,(\ref{W0}) should hold.
It was checked \cite{H1}  that Eq.\,(\ref{W0}) is satisfied for the nucleon without the electroweak 
structure.
Actually, Eq.\,(\ref{W0}) conforms to what we have got, with the difference that in our model, one observes explicitly the
validity of an analogous condition [to the end, the right hand side of our Eq.\,(\ref{pMnjl}) contains only the
pion production amplitude] 
also for the nucleons having the electroweak structure. In order to get sensible results beyond the soft pion limit,
the same condition should hold also in the approach \cite{H1}.
Otherwise, the pion production amplitude, derived directly, would differ from the one obtained via the current--current
amplitudes. This would be a dubious result, since it would not be clear, which amplitude is correct.
In our opinion, absence of a transparent proof of Eq.\,(\ref{W0}) \cite{H1} for the nucleons with the electroweak structure
makes the whole Haberzettl's calculation doubtful. Let us note that such a proof can be done only if this
structure is introduced not phenomenologically, but at a microscopic level.
This is achieved here within the concept of the hidden local symmetry and respecting the local chiral 
$SU(2)_R\times SU(2)_L$ invariance.

Moreover, it is not true that the right hand side of Eq.\,(19) \cite{H1} depends on $F_A$ only via ${\cal W}_\lambda$, as stated
in the paragraph after Eq.\,(21).
In other words, fulfilling Eq.\,(\ref{W0}) does not mean the elimination of the dependence of the pion electroproduction
amplitude on $F_A$.
The point is that it is the dynamical content of the model, which dictates the form factor of 
the Kroll--Ruderman term. As we have seen in our study, if the model respects the local chiral symmetry,
then this form factor is $F_A(k^2)$ in the soft pion limit, as it should be, in order to be in accord 
with the current algebra and PCAC. Actually, this result should be valid for any model possessing 
the local chiral symmetry.
On the other hand, models, respecting the gauge invariance only, will provide the Kroll-Ruderman term retaining the form
factor $F^V_1(k^2)$. So the claim \cite{H1} that the pion electroproduction processes at threshold cannot be used
to extract any information regarding the nucleon weak axial form factor should be considered as precipitous. 

On the contrary, the measurement of the nucleon weak axial form factor  $F_A(k^2)$ 
in the electroproduction of charged pions on the proton at threshold makes a good sense. 
In the recent measurement \cite{Lal} of this form factor by the $p(e,e'\,\pi^+)n$ reaction, the pion 
electroproduction amplitude Eq.\,(\ref{sppa}) with the added $\Delta$ excitation terms was used to
analyse the data. The value $M_A=(1.077\pm 0.039)GeV$ was found for the axial mass entering $F_A(k^2)$,
which is consistent with the value of $M_A$ known from neutrino scattering experiments.

Let us note that the study of the electroproduction of charged pions on the proton at threshold can also provide
the information on the pion charge radius \cite{BKM2},\cite{TN},\cite{EPT}.

\section*{Acknowledgments}

This work is supported by the grant GA \v{C}R 202/00/1669.  We thank Prof.~F.~C.~Khanna for
the critical reading of the manuscript and Mgr.~J.~Smejkal for discussions.

\end{document}